\documentclass[]{aspm}
\articleinfo{}{}{}

\usepackage{verbatim}
\usepackage{amssymb}
\usepackage{amsbsy}
\usepackage{amscd}
\usepackage{amsmath}
\usepackage{amsthm}
\usepackage[mathscr]{eucal}

\usepackage{epsfig,epsf}

\def\dd{\partial}
\def\p#1{(\ref{#1})}
\def\be{\begin{equation}}
\def\lb#1{\label{#1}}
\def\ee{\end{equation}}
\newfont{\bbd}{msbm10 scaled\magstep1}

\def\mL{\mathrm{L}}

\numberwithin{defn}{section}

\title[The hyperbolic modular double
and the Yang-Baxter equation]{The hyperbolic modular double and the Yang-Baxter equation}

\author[D. Chicherin, V. P. Spiridonov]{Dmitry Chicherin, Vyacheslav P. Spiridonov}

\address{Laboratoire d'Annecy-le-Vieux de Physique Th\'{e}orique, UMR 5108, Universit\'{e} de Savoie, CNRS,
B.P. 110,  F-74941 Annecy-le-Vieux, France; \\[0.1cm] Laboratory of Theoretical Physics, JINR, Dubna, Moscow region, 141980, Russia}

\email{chicherin@lapth.cnrs.fr; spiridon@theor.jinr.ru}

\rcvdate{}
\rvsdate{}

 \subjclass[2010]{AMS MSC}


\begin{document}

\begin{abstract}
We construct a hyperbolic modular double -- an algebra lying in between the Faddeev
modular double for $U_q(sl_2)$ and the elliptic modular double.
The intertwining operator for this algebra leads to an integral operator solution
of the Yang-Baxter equation associated with a generalized Faddeev-Volkov
lattice model introduced by the second author. We describe also the L-operator
and finite-dimensional R-matrices for this model.
\end{abstract}

\maketitle

\section{Introduction}

The representation theory is intimately related to special functions.
The quantum groups and Yang-Baxter equation (YBE) provide a wide class of novel
functions that do not appear in the classical representation theory of Lie groups.
These functions possess a number of peculiar properties and satisfy many intricate
identities which do not have classical counterparts.
The noncompact (or modular) quantum dilogarithm
\cite{Fad94,F95} is a remarkable special function  significant for a large class
of quantum integrable systems. In particular, it plays a prominent role in the
space-time discretization of the Liouville model and in the construction of
the lattice Virasoro algebra \cite{FKV,VF1}, as well as in the investigations
of the XXZ spin chain model in a particular regime \cite{jm,BT06}.

The observation that there exist two mutually commuting Weyl pairs
led Faddeev \cite{fad:mod} to the notion of a modular double of the quantum algebra $U_q(s\ell_2)$.
It is formed by two copies of $U_q(s\ell_2)$ with different deformation parameters
whose generators mutually (anti)commute with each other.
This doubling enables unambiguous fixing of the representation space of the algebra.

The elliptic modular double introduced by the second author in \cite{AA2008} carry over
the idea of doubling to the Sklyanin algebra \cite{skl1}. This doubling is extremely useful.
The symmetry constraints with respect to the extended algebra are much more powerful
as compared to the initial algebra. They enable again unambiguous description of the relevant objects.

The Faddeev-Volkov model \cite{VF} is a solvable two-dimensional lattice model of statistical
mechanics \cite{Baxter}.  In contrast to the Ising model,
its spin variables take continuous values. The Boltzmann weights are expressed in terms of
the modular quantum dilogarithm. In \cite{BMS} the free energy per edge
of this model was derived in the thermodynamic limit
using a particular form of the star-triangle relation.

A generalization of the Faddeev-Volkov model has been proposed by the second author in \cite{spi:conm}.
In this extension the Boltzmann weights are expressed in terms of the modular quantum dilogarithm as well,
but they have more involved form as compared to the original model.
The corresponding star-triangle relation is a degeneration of the elliptic beta integral
evaluation formula \cite{spi:umn}. The star-triangle relation associated with the
latter integral appeared first in the operator form as main identity
behind the integral Bailey lemma discovered in \cite{spi:bailey} (see also \cite{DS}
for a detailed discussion) and later it was formulated in the functional form in \cite{BS10}.

In the present work we study an algebraic structure
underlying the generalized Faddeev-Volkov model of \cite{spi:conm} and related quantum integrable systems.
First we consider a contraction of the Sklyanin algebra described in \cite{Gor93},
which is more general than $U_q(sl_2)$.
Then we show that this symmetry algebra can be enhanced using the doubling construction.
So, we will supplement the algebra with a dual set of generators (anti)commuting with
the initial generators. We baptize the resulting algebra as {\em the hyperbolic modular
double}, following the terminology of \cite{Ru} for the modular quantum dilogarithm
considered as a generalization of the Euler gamma function. It lies in between
the elliptic modular double and the modular double of $U_q(s\ell_2)$ in the sense that
the three algebras are arranged in a sequence of contractions.
We will pass naturally from the language of lattice models to the standard
YBE and find the most general solution of YBE having the symmetry of the hyperbolic modular double.
It is an integral operator acting on a pair of infinite-dimensional spaces which are representation
spaces of the latter algebra. An integral operator solution of the YBE
(at the plain non-deformed level)
was constructed for the first time in \cite{DKK01}. The factorization property of
the corresponding R-operator was noticed later in \cite{SD},
which resulted in a powerful almost purely algebraic techniques of
building general R-operators \cite{CD14,DKK07,DM,DS}.

Previously in \cite{CDS1,CDS} we described finite-dimensional solutions of YBE
for the elliptic modular double, modular double of $U_q(s\ell_2)$ and
the Lie group $\mathrm{SL}(2,\mathbb{C})$ in a concise form and elucidated
their factorization property \cite{CD15}.
One of the principal aims of this paper is to find all finite-dimensional solutions of YBE
having the symmetry of the hyperbolic modular double with generic deformation parameter.
However, a detailed construction of the representation theory of the latter algebra
together with the analysis of special functions associated with that, following
the considerations of \cite{NM} or  \cite{Gor93} and references therein, is left aside.
We do not discuss modular doubling of quantum
affine algebras as well, this subject was considered recently in \cite{MT}.

The plan of the paper is as follows. We start in Sect.~\ref{latMod} with a description
of the hyperbolic gamma function (modular quantum dilogarithm) and state its basic properties.
Then we present the solvable model of statistical mechanics generalizing the Faddeev-Volkov model
and the corresponding hyperbolic star-triangle relation.
In Sect.~\ref{latMod->R} we construct an integral R-operator in terms of the Boltzmann weights of this solvable vertex model and show that it satisfies the Yang-Baxter equation. We also rewrite it in the factorized form.
In Sect.~\ref{degSkl} we describe an algebra emerging as a degeneration of the Sklyanin algebra and construct
the corresponding intertwining operator of equivalent representations.
As a natural extension of this quantum algebra we introduce the hyperbolic modular double.
Then we study finite-dimensional irreducible representations of the hyperbolic modular double.
In Sect.~\ref{redR} we reduce the integral R-operator to a finite-dimensional
invariant subspace in one of its infinite-dimensional spaces.
In this way we obtain an explicit formula for finite-dimensional solutions of the YBE
with the symmetry of the hyperbolic modular double.
In Sect.~\ref{factL} we apply the reduction formula in the
simplest nontrivial setting. We choose the fundamental representation in one of the spaces
and recover the L-operator from the integral R-operator. It automatically takes the factorized form.
Finally, in Sect.~\ref{allFact} the reduction formula is elaborated further on
by a simplification to finite-dimensional matrix solutions of YBE such that all of
them take a factorized form.

\section{A solvable lattice model}
\label{latMod}

Gamma functions are the main building blocks in the construction of special
functions of hypergeometric type. The hierarchy of hyperbolic gamma functions
is formed by particular combinations of two multiple Barnes gamma functions
 \cite{bar} (the standard Jackson's $q$-gamma function \cite{aar}
is a combination of two Barnes gamma functions of the second order as well,
but we do not consider this function here).
The hyperbolic gamma function of the second order is a homogeneous function
of $u,\omega_1,\omega_2\in{\mathbb{C}}$. For $\text{Re}(\omega_1), \text{Re}(\omega_2)>0$ and
$0<\text{Re}(u)<\text{Re}(\omega_1+\omega_2)$ it has the form
\begin{align}
& \gamma^{(2)}(u;\omega_1,\omega_2):=
\exp\left(-\frac{\pi \textup{i}}{2}B_{2,2}(u;\omega_1,\omega_2)- \right. \notag \\ & \makebox[5em]{} \left.
-\int_{\mathbb{R}+\textup{i} 0}\frac{e^{ux}}
{(1-e^{\omega_1 x})(1-e^{\omega_2 x})}\frac{dx}{x}\right),
\label{hgf}
\end{align}
where $B_{2,2}$ is a multiple Bernoulli polynomial of the second order
$$
B_{2,2}(u;\omega_1,\omega_2)=\frac{1}{\omega_1\omega_2}
\left( \left(u-\frac{\omega_1+\omega_2}{2}\right)^2
-\frac{\omega_1^2+\omega_2^2}{12}\right).
$$
Denoting $q=e^{2\pi \textup{i}\omega_1/\omega_2}$ and
$\tilde q=e^{-2\pi \textup{i}\omega_2/\omega_1}$ and
assuming that $|q|<1$, one can write
$$
\exp\left(-\int_{\mathbb{R}+\textup{i} 0}\frac{e^{ux}}
{(1-e^{\omega_1 x})(1-e^{\omega_2 x})}\frac{dx}{x}\right)
=\frac{(e^{2\pi \textup{i} u/\omega_1}\tilde q; \tilde q)_\infty}
{(e^{2\pi \textup{i}  u/\omega_2}; q)_\infty},
$$
where $(t;q)_\infty=\prod_{k=0}^\infty(1-tq^k)$.

The modular quantum dilogarithm is usually defined as a function
obtained from \eqref{hgf} by removing the exponential factor involving
$B_{2,2}(u;\omega_1,\omega_2)$, the shift $u\to u+(\omega_1+\omega_2)/2$,
and some renormalization of the variables $u$ and $\omega_{1,2}$.
In particular, in the context of $2d$ conformal field theory it is accepted
to denote $\omega_1=b,\, \omega_2=b^{-1}$.
We shall use the following representation
\begin{align}
& \gamma(z):=\gamma(z;b) :=
\exp \left(- \frac{\textup{i} \pi}{2}\left(z-\frac{b+b^{-1}}{2}\right)^2 +
\right. \notag \\
& \makebox[5em]{} \left. + \frac{\textup{i} \pi}{24}(b^2+b^{-2}) + \int\limits^{+\infty}_{-\infty} \frac{d t}{4t} \frac{e^{t(2z-b-b^{-1})}}{\sin( \textup{i} b t) \sin( \textup{i} b^{-1} t)}
\right),\lb{hypg}
\end{align}
where the integration contour goes above the singularity at $t$ = 0.
One can easily restore the original function
$$
\gamma^{(2)}(z;\omega_1,\omega_2) = \gamma(z/\sqrt{\omega_1\omega_2};\sqrt{\omega_1/\omega_2})\,.
$$
This integral representation is valid for ${\rm Re}(b) > 0$ and $0<{\rm Re}(z)<{\rm Re}(b+b^{-1})$.
The analytic continuation enables one to extend the definition \p{hypg}
to a wider range of parameters.

The inverse of this special function is called
also the double sine-function and denoted either $S(u;\omega_1,\omega_2)$
\cite{jm} or $S_b(z)$ \cite{BT06}. The notation $\gamma^{(2)}(z;\omega_1,\omega_2)$
is taken from \cite{spi:conm}.
Here we use the terminology suggested in \cite{Ru}.
Interrelations between main known modifications
of function \p{hypg} are described in Appendix A of \cite{spi:conm}.
Various identities for the quantum dilogarithm can be found
in \cite{FKV,Vol05} and some other papers.

The definition \p{hypg} implies that $\gamma(z)$ is invariant under the swap
$b \leftrightarrows b^{-1}$. It satisfies two linear difference equations of
the first order
\begin{align} \lb{period}
\gamma(z+b) = 2 \sin (\pi b z)\, \gamma(z) ,\qquad
\gamma(z+b^{-1}) = 2 \sin (\pi b^{-1} z)\, \gamma(z).
\end{align}
Let us introduce the crossing parameter
$$
\eta := -\frac{b+b^{-1}}{2}.
$$
Then the reflection formula can be written as follows
\begin{align} \label{refl}
\gamma(z)\,\gamma(-2\eta-z) = 1.
\end{align}
The function $\gamma(z)$ is meromorphic. It has a double series of zeros
\begin{align} \lb{zero}
z =  b (n+1) +  b^{-1} (m+1) ,\;\;\; n,\,m \in \mathbb{Z}_{\geq 0},
\end{align}
and a double series of poles
\begin{align} \lb{pole}
z = - b \,n - b^{-1} m ,\;\;\; n,\,m \in \mathbb{Z}_{\geq 0}.
\end{align}
In the following we deal with multiple products of the hyperbolic gamma function.
In order to avoid lengthy formulae we adopt the convention
$$
\gamma(\pm x + g) := \gamma(x + g) \,\gamma(-x + g)\,,\;\;
\gamma(\pm x \pm y) := \gamma(\pm x + y) \,\gamma(\pm x - y).
$$

In \cite{spi:conm} a nontrivial generalization of the Faddeev-Volkov model
\cite{VF1,VF} has been proposed.
Such models of statistical mechanics are defined on the square lattice.
The continuous spin variables sit in the lattice vertices.
The rapidity variables are associated with the medial graph built
with the help of pairs of directed lines crossing edges in the middle
with the inclination of 45 or 135 degrees (see, e.g., a picture on Fig. \ref{fig2}).
Self-interaction of spins and interactions between the nearest neighboring spins are allowed.
The Boltzmann weight $W(\alpha-\beta; x,y)$ is assigned to a horizontal edge connecting
a pair of vertices with spins $x$, $y$ that is crossed by a pair of
medial graph lines carrying the rapidities $\alpha$ and $\beta$.
Similarly the Boltzmann weight $\overline{W}(\alpha-\beta; x,y)$ is assigned to a
vertical edge connecting a pair of vertices with the spins $x$, $y$ that is crossed by a
pair of medial graph lines carrying rapidities $\alpha$ and $\beta$.
The self-interaction at a vertex $z$ contributes the Boltzmann weight $\rho(z)$ to the partition function.
The edge Boltzmann weights $W$ and $\overline{W}$ of the model \cite{spi:conm}
are given by fourfold products of hyperbolic gamma functions
and the vertex Boltzmann weight $\rho$ is a product of two hyperbolic gamma functions
\begin{align} \lb{BW}
& W(\alpha;x,y) = \gamma(\alpha-\eta \pm \textup{i} x \pm \textup{i} y) \,, \\ \notag &
\overline{W}(\alpha;x,y) = \gamma(-\alpha \pm \textup{i} x \pm \textup{i} y) \,,\;\;\;
\rho(z) = \frac{1}{2\gamma(\pm 2 \textup{i} z)}\,.
\end{align}
The edge Boltzmann weights depend on the difference of rapidities and they
are symmetric in the spin variables $W(\alpha;x,y) = W(\alpha;y,x)$,\\ $\overline{W}(\alpha;x,y) = \overline{W}(\alpha;y,x)$.
Let us recall that the Boltzmann weights depend on $b$ which is a temperature-like parameter.

Physical interpretation of $W$ and $\overline{W}$ as true Boltzmann weights
requires their positivity. This is possible in two regimes of the key parameter $b$:
1) $b$ is real and $0<b<1$; 2) $|b|=1,$ Im$(b^2)>0$. In both regimes $\eta<0$
and $\gamma(z)^*=\gamma(z^*)$. As a result one should demand that the spin variables
are real. As to the rapidities, one can set $\eta < \alpha <-\eta$ for $W(\alpha)$
and $0<-\alpha <-2\eta$ for $\overline{W}(\alpha)$.
These constraints should correspond to unitary representations of the
hyperbolic modular double to be described below.

The Boltzmann weights possess a crossing symmetry, i.e. the horizontal and vertical
edge weights are related as follows
\begin{align} \lb{crossing}
\overline{W}(\alpha;x,y) = W(\eta-\alpha;x,y).
\end{align}
We note that in view of the reflection formula \p{refl} and quasiperiodicity \p{period}
the vertex Boltzmann weight $\rho(z)$ can be rewritten solely in terms of
the trigonometric functions, i.e. its expression in terms of the hyperbolic gamma
functions is overcomplicated. In contrast, the edge Boltzmann weights $W$ and $\overline{W}$
are genuine products of hyperbolic gamma functions and the number of such
functions in the products cannot be reduced.

\begin{figure}[h]
$$
\begin{array}{c}
\includegraphics[width = 4.0 cm]{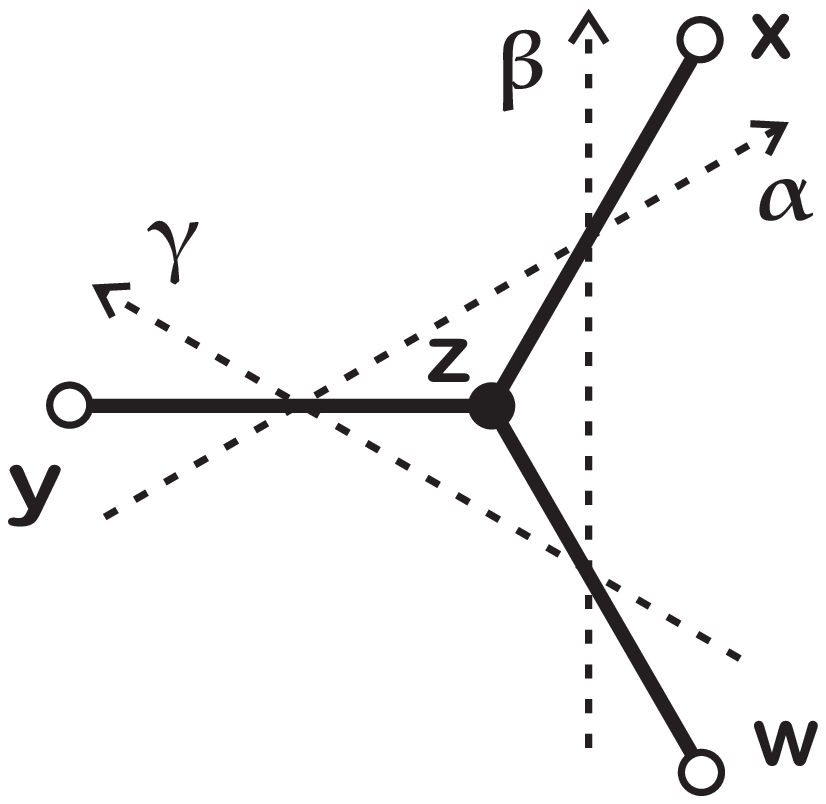}
\end{array} \quad = \quad \chi \;\; \times \quad
\begin{array}{c}\includegraphics[width = 4.0 cm]{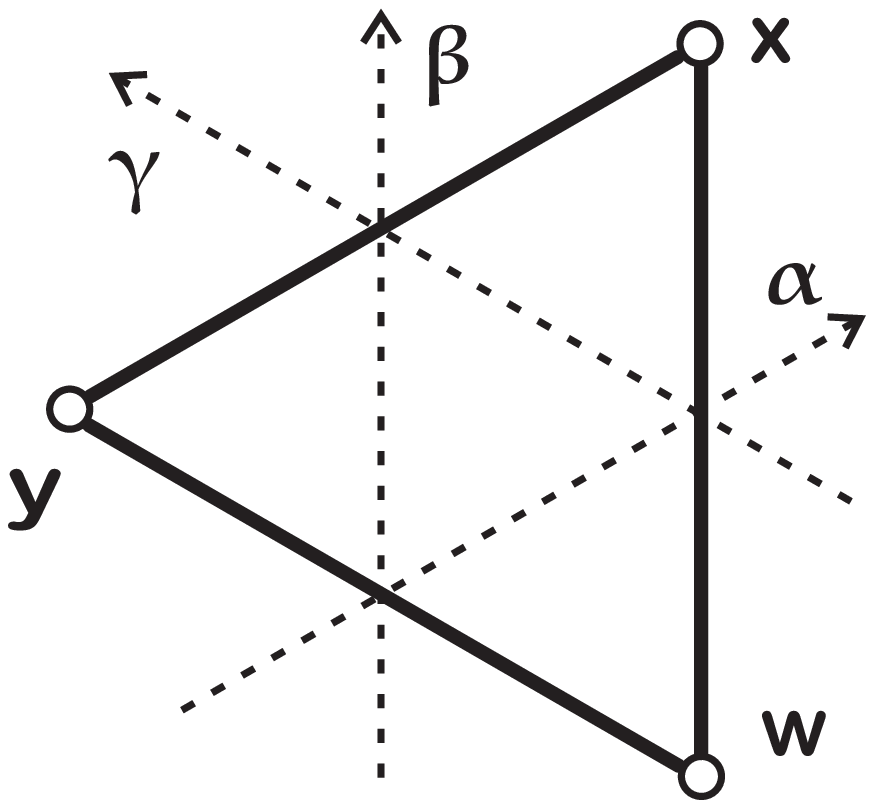}\end{array}
$$
\caption{\label{f1} The star-triangle relation.}
\end{figure}

The formulated model is solvable because of the star-triangle relation depicted in Fig. \ref{f1}.
This relation equates the partition functions of two elementary cells,
\begin{align} \label{HSTR}
\int\limits^{+\infty}_{-\infty} \rho(z) \, \overline{W}(\alpha-\beta;x,z)\, W(\alpha - \gamma;y,z)\, \overline{W}(\beta-\gamma;w,z) \,d z \notag \\ =
\chi(\alpha,\beta,\gamma)\,W(\alpha-\beta;y,w)\,\overline{W}(\alpha-\gamma;x,w) \,
W(\beta-\gamma;x,y),
\end{align}
up to a normalization constant $\chi$. Using this example one can see that
the horizontal edge Boltzmann weights $W(\alpha-\beta; x,y)$ depends on the difference $\alpha-\beta$,
where $\alpha$  is the rapidity of the upward directed median line of 45 degrees
and $\beta$ is the rapidity of the upward directed median line of 135 degrees.
For the vertical edge weights $\overline{W}(\alpha-\beta; x,y)$ the situation is similar --
$\alpha$ is the rapidity of the line going to the right of the edge and $\beta$ -- of
the line going to the left.
We call identity \p{HSTR} with the weights \p{BW} the hyperbolic star-triangle relation.
Corresponding normalization constant has the following form
$$
\chi(\alpha,\beta,\gamma) = \gamma\left(2\beta-2\alpha\right)\gamma\left(2\gamma-2\beta\right)
\gamma\left(2\alpha-2\gamma-2\eta\right).
$$
This local relation enables one to construct a family of commuting
row-to-row transfer matrices and then to
calculate the partition function of the model using the machinery of QISM \cite{Baxter}.
The free energy per edge of the model in the thermodynamical limit
has been calculated in \cite{spi:conm} following the method from \cite{BMS,BS10}.

Let us substitute in \eqref{HSTR}
\begin{equation}
W(\alpha;x,y)= m(\alpha) W_r(\alpha;x,y), \;\;
\overline{W}(\alpha;x,y)=  m(\eta-\alpha)\overline{W}_r(\alpha;x,y)
\label{reno}\end{equation}
and choose the function $m(\alpha)$ in such a way that the normalization constant
$\chi$ on the right-hand side of \eqref{HSTR} disappears,
$$
\frac{m(\alpha-\beta)m(\eta-\alpha+\gamma)m(\beta-\gamma)}
{m(\eta-\alpha+\beta)m(\alpha-\gamma)m(\eta-\beta+\gamma)}\, \chi(\alpha,\beta,\gamma) =1.
$$
Ascribe now to the edges the renormalized weights $W_r(\alpha;x,y)$
and \\ $\overline{W}_r(\alpha;x,y)$. Then, denoting the total number of edges in the infinitely
growing lattice as $N$, one finds that the free energy per edge
$$
\beta f_{edge} = - \underset{ N\to \infty}{\lim} \frac{\log Z^{(r)}}{N} =0,
$$
where $Z^{(r)}$ is the total partition function for the model with
renormalized Boltzmann weights.

Equivalently, we can keep the original Boltzmann weights intact and
compute the contribution of the renormalizing factors in the asymptotics
of the partition function. Take the finite rectangular lattice with $N$ spins
along the horizontal
line and $M$ spins along the vertical line. Such lattice has $(N-1)M$ horizontal
edges and $N(M-1)$ vertical edges.  Therefore the indicated renormalization of
the Boltzmann weights yields a scaling of the partition function
$$
Z_{N,M}=  Z_{N,M}^{(r)}\, m(\alpha)^{(N-1)M}\, m(\eta-\alpha)^{N(M-1)},
$$
i.e. the free energy per edge of the original models is
$$
\text{ free energy per edge}= - \underset{ N,M\to \infty}{\lim} \frac{\log Z_{N,M}}{NM}=- \log m(\alpha)m(\eta-\alpha).
$$

It is easy to see that the needed normalization constant $m(\alpha)$ is found
from the equation
\be
m(\alpha+\eta)=\gamma(2\alpha;b)m(-\alpha).
\label{modeqm}\ee
As shown in \cite{spi:conm}, the solution of this equation satisfying
the unitarity relation $m(\alpha)m(-\alpha)=1$
is given by the ratio of two hyperbolic gamma functions of the third
order $\gamma^{(3)}(u;\omega_1,\omega_2,\omega_3)$ for a special choice
of the quasiperiods $\omega_i$. In the current notation it has the  following
integral representation
\begin{eqnarray} \nonumber &&
m(\alpha)=
\exp\Big(- \pi \textup{i}\left(\alpha^2+\frac{1}{24}(1-2(b+b^{-1})^2)\right)+
\\ && \makebox[4em]{}
+\frac{1}{8}\int_{\mathbb{R}+\textup{i}0}\frac{e^{4\alpha t}}{\sin(\textup{i}bt) \sin(\textup{i}b^{-1}t)
\cos(\textup{i}(b+b^{-1})t)}
\frac{dt}{t}\Big).
\label{m}\end{eqnarray}
It happens that this result coincides with a similar normalization
factor $m(\alpha)$ for the
original Faddeev-Volkov model derived in \cite{BMS}.

The hyperbolic star-triangle relation \p{HSTR} can be written as the integral identity
\begin{align} \label{hypgamma}
\int\limits^{+\infty}_{-\infty} \prod_{k = 1}^{6} \gamma(g_k \pm \textup{i} z) \frac{d z}{2\gamma(\pm 2 \textup{i} z)}
= \prod_{1 \leq j < k \leq 6}\gamma(g_j + g_k),
\end{align}
where parameters $g_k,\, k = 1 , \ldots,6,$ satisfy the
constraints Re$(g_k)>0$ and the balancing condition
$$
\sum_{k = 1}^{6} g_k = -2 \eta.
$$
Note that the condition Re$(g_k)>0$ restricts the
domain of complex values of the rapidities and spins as follows
$$
\text{Re}(\beta-\alpha\pm\textup{i} x), \;
\text{Re}(\alpha-\gamma-\eta\pm\textup{i} y), \;
\text{Re}(\gamma-\beta\pm\textup{i} w) >0.
$$
In principle these restrictions can be relaxed by the analytical continuation
(e.g., it can be reached by a deformation of the integration contour).

The first mathematically rigorous proof of relation \eqref{hypgamma}, and, so,
of the star-triangle relation \eqref{HSTR}, was obtained in  \cite{St}.
However, this identity is a special limiting
case of the elliptic beta integral evaluation formula rigorously
established in \cite{spi:umn}.
Let us note that the exactly computable integral \p{hypgamma} is a non-compact
(hyperbolic) analogue of the Rahman (trigonometric) $q$-beta integral \cite{Rah}.


Another important property of the described Boltzmann weights is the unitarity relation.
For real values of the spins $x$ and $y$ one has
\begin{align}
&\int\limits^{+\infty}_{-\infty} \rho(z)\, \overline{W}(\alpha-\beta;x,z)\,
\overline{W}(\beta-\alpha;y,z)\, d z \notag \\
&\kern 35pt = \frac{1}{2\rho(x)} \,\gamma(2\alpha-2\beta)\gamma(2\beta-2\alpha)
\left( \delta(x-y) + \delta(x+y) \right).
 \lb{unit}
\end{align}
This identity can be rigorously obtained by taking the limit $\gamma \to \alpha$
in \p{HSTR}. In the computation of partition functions of three-dimensional
supersymmetric field theories on the squashed spheres such relations indicate
the chiral symmetry breaking phenomenon \cite{SV2014}.
The Boltzmann weights $W$ and $\overline{W}$ satisfy the reflection equation that
is an evident consequence of Eq.~\p{refl},
\begin{align} \lb{reflW}
W(\alpha-\beta;x,y) \,W(\beta-\alpha;x,y) = 1.
\end{align}

\section{From the lattice model to the integral R-operator}
\label{latMod->R}

The star-triangle relations imply integrability of the two-dimensional lattice
models, similar to the case outlined in the previous section. Another wide class
of integrable systems is associated with
the quantum spin chains. Formulation of the latter models in the framework of QISM \cite{Fad96,KS81}
requires definition of the R-matrix solving the YBE.
In this section we show how to construct such R-matrices associated with the hyperbolic star-triangle
relation \p{HSTR}. Since the spin variables sitting in vertices of the two-dimensional lattice
take continuous values (in contrast to the discrete spins of the Ising and chiral Potts models),
on the spin chains side we deal with integral operators instead of the finite-dimensional R-matrices.
In other words, the quantum spaces of the relevant spin chain are infinite-dimensional functional spaces.
Therefore we call solutions of the corresponding YBE the R-operators to emphasize this aspect.
We will indicate in Sect.~\ref{redR} that the integral R-operators represent in some sense the most general
YBE solutions, since they embrace all finite-dimensional R-matrices.
Our presentation below follows to some extent the original construction of \cite{DKM}.

Thus we are interested in the integral operator $\mathbb{R}_{12}(u|g_1,g_2)$
defined on the tensor product of two infinite-dimensional function spaces that are representation spaces
with labels (``spins") $g_1$ and $g_2$ (arbitrary complex numbers) of some algebra.
The symmetry algebra underlying the hyperbolic star-triangle relation will be introduced
in Sect.~\ref{degSkl}. The R-operator depends on a complex number $u$ -- the spectral parameter and
satisfies the YBE
\begin{align}
&\mathbb{R}_{12}(u-v|g_1,g_2) \,\mathbb{R}_{13}(u|g_1,g_3) \,\mathbb{R}_{23}(v|g_2,g_3) \notag\\ &=
\mathbb{R}_{23}(v|g_2,g_3) \,\mathbb{R}_{13}(u|g_1,g_3)\, \mathbb{R}_{12}(u-v|g_1,g_2).\label{YBE}
\end{align}

\begin{figure}[h]
\begin{align*}
&\bigl[\mathbb{R}_{12}(u_1,u_2|v_1,v_2)\, \Phi\bigr](z_1,z_2)= \\[0.2cm] & =
\int\limits^{+\infty}_{-\infty}\int\limits^{+\infty}_{-\infty}  \quad \begin{array}{c}\includegraphics[width = 4 cm]{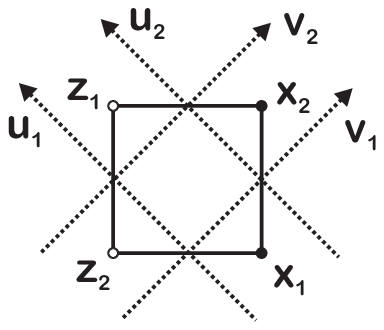}\end{array}
\Phi(x_1,x_2) \, d x_1 \, d x_2
\end{align*}
\caption{\label{fig2}The kernel of the integral R-operator is the partition function of an elementary square cell.}
\end{figure}

Instead of the integral operator $\mathbb{R}_{12}(u|g_1,g_2)$, we can
consider first a more general notation operator $\mathbb{R}_{12}(u_1,u_2|v_1,v_2)$
depending on four complex parameters and satisfying the equation
\begin{align} \lb{YBErap}
&\mathbb{R}_{12}(u_1,u_2|v_1,v_2) \,\mathbb{R}_{13}(u_1,u_2|w_1,w_2) \,\mathbb{R}_{23}(v_1,v_2|w_1,w_2) \notag \\
&=\mathbb{R}_{23}(v_1,v_2|w_1,w_2) \,\mathbb{R}_{13}(u_1,u_2|w_1,w_2)\, \mathbb{R}_{12}(u_1,u_2|v_1,v_2).
\end{align}
This equation is rather similar to \p{YBE}.  An operator solution of Eq.~\p{YBErap} can be easily constructed
in terms of the lattice model formulated in the previous section. The parameters $u_1,u_2,v_1,v_2,w_1,w_2$
are identified with the rapidities. The kernel of the integral operator is a product of four edge Boltzmann
weights (two horizontal and two vertical) and two vertex Boltzmann weights,
\begin{align} \label{Rint}
&\bigl[\mathbb{R}_{12}(u_1,u_2|v_1,v_2)\Phi\bigr](z_1,z_2)  =
\int\limits^{+\infty}_{-\infty}\int\limits^{+\infty}_{-\infty} \rho(x_1)\rho(x_2) W(u_1-v_2;z_1,z_2)\times  \\ &
\overline{W}(u_1-v_1;z_1,x_2)
\overline{W}(u_2-v_2;z_2,x_1)
W(u_2-v_1;x_1,x_2) \Phi(x_1,x_2) d x_1 d x_2. \notag
\end{align}
Thus the kernel of $\mathbb{R}_{12}(u_1,u_2|v_1,v_2)$
is the partition function of an elementary square cell, see Fig. \ref{fig2}.

Taking into account explicit expressions for the Boltzmann weights we rewrite Eq.~\p{Rint}
as follows
\begin{align}
&\left[\mathbb{R}_{12}(u_1,u_2|v_1,v_2)\Phi\right](z_1,z_2)
= \int\limits^{+\infty}_{-\infty}\int\limits^{+\infty}_{-\infty}
\frac{d x_1 \,d x_2\,\Phi(x_1,x_2)}{4\gamma(\pm 2 \textup{i} x_1)\gamma(\pm 2 \textup{i} x_2)} \times \notag \\
& \makebox[4em]{} \gamma(\pm \textup{i} z_1 \pm \textup{i} z_2 + u_1 -v_2 -\eta)\,
\gamma(\pm \textup{i} x_1 \pm \textup{i} z_2 + v_1-u_1) \times \notag \\
& \makebox[4em]{} \gamma(\pm \textup{i} x_2 \pm \textup{i} z_1 + v_2-u_2)\,\gamma(\pm \textup{i} x_1 \pm \textup{i} x_2 + u_2 - v_1 - \eta).
\lb{explR}
\end{align}
This R-operator corresponds to the generalized Faddeev-Volkov model of \cite{spi:conm}.
It can be derived from the elliptic hypergeometric R-operator constructed in \cite{DS} after taking
a particular limit in the parameters, but the rigorous proof of this fact would
require quite intricate techniques. Remind that the latter  R-operator
intertwines representations of the Sklyanin algebra.

Let us recall that the vertex Boltzmann weight $\rho$ is in fact a trigonometric function.
The integrand function in the expression \p{explR} is a genuine product of 16 hyperbolic gamma functions (modulo a trigonometric multiplier) and their number cannot be reduced.
In \cite{CD14} the R-operator associated with the Faddeev modular double of $U_q(s\ell_2)$
has been constructed in a similar form. It corresponds to the Faddeev-Volkov lattice model itself.
However, the integrand function of the corresponding integral operator solution of YBE is a product
of only 8 hyperbolic gamma functions (up to a trigonometric multiplier).
One can obtain this R-operator rigorously from expression \p{explR} by taking appropriate limits
in the parameters such that the $\gamma$-functions depending on the sum $x_1 + x_2$ disappear.
E.g., such a limiting procedure is described in \cite{spi:conm} for the reduction of
the hyperbolic star-triangle relation down to the original Faddeev-Volkov model case.

\begin{figure}[h] \begin{center}\includegraphics[width = 10cm]{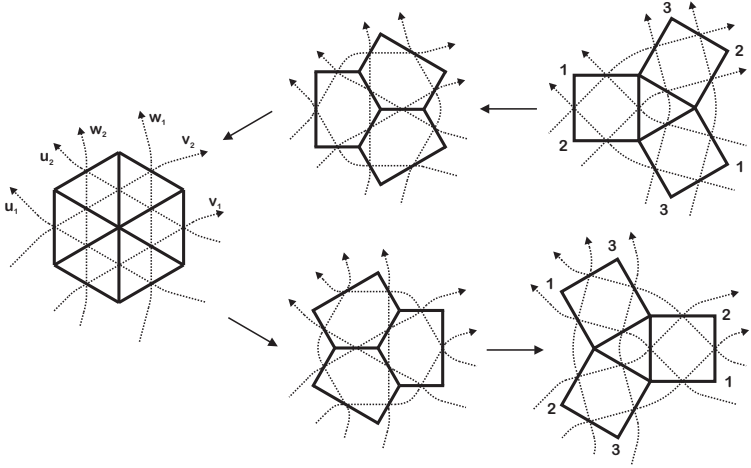}\end{center}
\caption{\label{fig3}The sequence of star-triangle moves transforming the left-hand side expression in
\p{YBErap} to its right-hand form.}
\end{figure}

Similar to the very first integral R-operator considerations in \cite{DKM},
it is easy to check that the operator $\mathbb{R}_{12}(u_1,u_2|v_1,v_2)$, Eq. \p{Rint},
solves Eq.~\p{YBErap}.
In order to demonstrate how it works we resort to graphical representation of
the integral R-operators. In Fig. \ref{fig3} at the right top we depicted convolution
of the kernels of the integral operators from
the left-hand side expression in Eq.~\p{YBErap}
and at the right bottom we depicted convolution of the kernels from the RHS of Eq.~\p{YBErap}.
External points are marked by numbers $1,2,3$. They denote three quantum spaces.
Integration over internal points is assumed (convolution of the kernels). The dotted lines
are the rapidity lines.
The left-hand and right-hand side expressions in Eq.~\p{YBErap} are connected to each other by a sequence of moves. Each move is the application of the star-triangle transformation, Eq.~\p{HSTR}.
Thus the YBE \p{YBErap} boils down to a combination of the star-triangle relations \p{HSTR}.
Keeping track of the normalization factors $\chi$ arising at each step
one can check that they eventually drop out.

However, we still need to understand the algebraic meaning of solution \p{Rint},
which is the primary goal of the present paper. In other words, we have to
give an algebraic interpretation of the rapidities $u_1,u_2,v_1,v_2$. As we will see further
they can be chosen as linear combinations of the spectral parameters $u, v$ and the
representation labels $g_1,g_2$ of a certain algebraic structure which is studied in the next section,
\begin{align} \label{uv}
u_1 =
\frac{u+g_1}{2}\ ,\
u_2 = \frac{u-g_1}{2}\ \ ,\ \
v_1 =
\frac{v+g_2}{2}\ , \
v_2 = \frac{v-g_2}{2}\ .
\end{align}
This relation yields a solution of the YBE in the form~\p{YBE},
$$
\mathbb{R}_{12}(u-v|g_1,g_2)= \mathbb{R}_{12}(u_1,u_2|v_1,v_2).
$$
Our considerations below indicate that this is in fact the most general solution of YBE
compatible with certain quantum algebra. The most strong argument follows from the
fact that Eq.~\p{Rint} embraces all finite-dimensional solutions of the YBE~\p{YBE}
associated with this algebra.

Factorization of the kernel of the integral operator \p{Rint} implies factorization of
 the operator itself. Indeed, it can be written as a product of five elementary operators
\begin{align} \lb{R4fact}
\mathbb{R}_{12}(u_1,u_2|v_1,v_2) = \mathrm{P}_{12}\,\mathrm{S}(u_1-v_2)\,\mathrm{M}_2(u_2-v_2)\,
\mathrm{M}_1(u_1-v_1)\,\mathrm{S}(u_2-v_1).
\end{align}
Here $\mathrm{P}_{12}$ is a permutation operator of two tensor factors, i.e. $\mathrm{P}_{12}\Phi(z_1,z_2)\\ = \Phi(z_2,z_1)$.
$\mathrm{S}(u)$ is an operator of multiplication by a particular function,
\begin{align} \lb{S}
\mathrm{S}(u) = W(u;z_1,z_2)= \gamma(\pm \textup{i} z_1 \pm \textup{i} z_2 + u -\eta).
\end{align}
$\mathrm{M}_1$ and $\mathrm{M}_2$ are two copies of the integral operator
\begin{align} \lb{Mbw}
\bigl[\mathrm{M}(g)\,\Phi \bigr](z) = \frac{1}{\gamma(-2g)}\int\limits^{+\infty}_{-\infty} \rho(x) \, \overline{W}(g;z,x) \,\Phi(x)\,d x
\end{align}
acting in the first and second quantum spaces, respectively.
This operator is a degeneration of an elliptic hypergeometric integral operator introduced in \cite{spi:bailey}.
Owing to the chosen definition \p{Mbw}, normalizations of the R-operators
\p{R4fact} and \p{explR} differ by the multiplicative numerical factor
$\gamma(2v_1-2u_1)\gamma(2v_2-2u_2)$. This renormalization of the R-operator
removes certain divergences appearing during the reduction we describe
below and makes this procedure smooth.

At this point we should specify an appropriate function space for the operator $\mathrm{M}$, Eq.~\p{Mbw}.
Firstly, the kernel of $\mathrm{M}$ is invariant with respect to the reflections
$x \to -x$ and $z \to -z$.
Consequently, $\mathrm{M}$ projects out odd functions and maps onto the space of even functions.
Moreover, calculating the asymptotics of the kernel at $x \to \pm \infty$ we obtain a restriction
on the asymptotic behavior of $\Phi(x)$. Thus we assume that $\Phi$ is an even function, i.e. $\Phi(-x) = \Phi(x)$,
and $e^{4\pi \textup{i} x g}\Phi(x)$ is rapidly decaying at $x \to +\infty$.

In view of the reflection equation~\p{reflW} and unitarity, Eq.~\p{unit}, of the Boltzmann weights
the operators $\mathrm{S}$ and $\mathrm{M}$ satisfy (in the space of even functions) the inversion relations
\begin{align} \lb{sq}
\mathrm{S}(u) \, \mathrm{S}(-u) = 1 ,\qquad \mathrm{M}(g)\, \mathrm{M}(-g) = 1.
\end{align}
The second relation is a degeneration of the inversion formula proved in \cite{SW} for the
elliptic hypergeometric integral operator of \cite{spi:bailey}.
The inversion relation \p{sq} for $\mathrm{S}(u)$ is valid for generic values of
$g \in \mathbb{C}$, but for $\mathrm{M}(g)$ it is violated for particular discrete
lattice points on ${\mathbb{C}}$. In the latter case a nontrivial null-space of $\mathrm{M}(g)$
appears which is described in the next section.
Let us note that for generic parameters the operators $\mathrm{S}$, $\mathrm{M}_1$, $\mathrm{M}_2$
provide a twisted representation of generators of the permutation group $S_4$ satisfying the Coxeter
relations. More precisely, the star-triangle relation \p{HSTR} can be reformulated as
the cubic Coxeter relations for $\mathrm{S}$, $\mathrm{M}_1$, $\mathrm{M}_2$,
whereas the inversion relations \eqref{sq} represent quadratic Coxeter relations.
For more details on this interpretation, see the end of Sect.~\ref{factL} and
\cite{SD,DKK07,DM,DS} where the allied constructions are elaborated in detail.
Finally we note that the identities \p{sq} lead to the unitarity-like relation for the R-operator \p{R4fact},
\begin{align}
\mathbb{R}_{12}(u|g_1,g_2)\,\mathbb{R}_{12}(-u|g_2,g_1) = 1.
\end{align}

\section{The intertwining operator of a degenerate Sklyanin algebra}
\label{degSkl}

In this section we study in detail the operator $\mathrm{M}$, Eq.~\p{Mbw}, in order to infer its algebraic meaning.
We rewrite Eq.~\p{Mbw} explicitly in terms of the hyperbolic gamma functions
\begin{align} \lb{M}
\bigl[\mathrm{M}(g)\,\Phi \bigr](z) = \frac{1}{2}\int\limits^{+\infty}_{-\infty}
\frac{\gamma(-g \pm \textup{i} z \pm \textup{i} x)}
{\gamma(\pm 2 \textup{i} x)\gamma(-2g)} \Phi(x)\,d x\, .
\end{align}

For certain discrete values of $g$ the operator $\mathrm{M}(g)$ simplifies considerably.
First of all at the origin $g=0$ it is the identity operator, $\mathrm{M}(0) =1$,
which can be seen by taking the limit $g\to 0$ after a simple residue calculus.
Moreover one can easily check that $\mathrm{M}(g)$ respects a pair of the
contiguous relations involving the shifts of $g$ by $\frac{b}{2}$ and $\frac{1}{2b}$,
\begin{eqnarray}\label{rec1}
&&
- \frac{\textup{i}}{\sin(2\pi \textup{i} b z)} \,\sin({\textstyle \frac{b}{2}\dd_z}) \,
 \mathrm{M}(g) = \textstyle \mathrm{M}(g+\frac{b}{2}) ,
\\ &&
- \frac{\textup{i}}{\sin(2\pi \textup{i} b^{-1} z)} \,\sin({\textstyle\frac{1}{2b}\dd_z}) \,
\mathrm{M}(g) = \textstyle \mathrm{M}(g+\frac{1}{2b}).
\label{rec2}\end{eqnarray}
Elliptic analogues of these relations can be found in \cite{CDKK12,DS2}.
Applying recurrences \eqref{rec1}, \eqref{rec2} for constructing $\mathrm{M}(g)$ in the
discrete quarter-infinite lattice of points $g=\frac{n b}{2}+\frac{m}{2b}$,
$n,\,m \in \mathbb{Z}_{\geq 0}$, we find that the integral operator \eqref{M}
is converted to a product of $n+m$ finite-difference operators of the first order
\begin{align} \lb{Mfact}
\mathrm{M}({\textstyle \frac{n b}{2}+\frac{m}{2b}}) =
\left[ \frac{-\textup{i}}{\sin(2\pi \textup{i} b z)} \,\sin({\textstyle \frac{b}{2}\dd_z}) \right]^{n}
\left[ \frac{-\textup{i}}{\sin(2\pi \textup{i} b^{-1} z)} \,\sin({\textstyle \frac{1}{2b}\dd_z}) \right]^{m}.
\end{align}

We have already mentioned above that the R-operator \p{Rint}
is related to a certain quantum algebra. Let us now specify it explicitly.
It is a contraction of the Sklyanin algebra~\cite{skl1} that has been introduced
in \cite{Gor93} and then investigated in \cite{Smir10,Ant97,Brad01}.
This degenerate Sklyanin algebra is formed by four generators $A,B,C,D$ which
respect the following commutation relations
\begin{align}
& C \,A = e^{\textup{i} \pi b^2} \,A\, C \, ,\;\;
D \,C = e^{\textup{i} \pi b^2} \,C\, D \,, \notag \\
& [\,A\,,\,D\,] = -2 \textup{i} \sin^3 \pi b^2 \, C^2\, ,\;\;
[\,B\,,\,C\,] = \frac{A^2-D^2}{2 \textup{i} \sin \pi b^2} \,, \notag \\
& A\,B - e^{\textup{i} \pi b^2}\, B\,A = e^{\textup{i} \pi b^2} \, D \, B -B\,D
= \frac{\textup{i}}{2} \sin 2 \pi b^2 \, (C\,A-D\,C)\,.\lb{alg}
\end{align}
It has a pair of Casimir operators
\begin{align}
&K_0 =  e^{\textup{i} \pi b^2} A\, D - \sin^2 \pi b^2  \, C^2 ,\notag\\
&K_1 = \frac{e^{-\textup{i} \pi b^2} A^2 + e^{\textup{i} \pi b^2} D^2}{4\sin^2 \pi b^2} -B\, C -\frac{1}{2}\cos \pi b^2 \, C^2 \label{Cas}
\end{align}
commuting with all four generators. This algebra is different from the conventional
quantum deformation of the rank 1 Lie algebra, $U_q(s\ell_2)$ \cite{Kul83}, in particular,
it does not obey the Hopf algebra structure. As shown in \cite{Gor93},
the spectral problem for a special quadratic combination of the generators
of this algebra reproduces the eigenvalue problem for the Askey-Wilson
polynomials. Therefore this algebra comprises the Zhedanov algebra
as well \cite{Z}, which was constructed precisely with the aim of
interpreting Askey-Wilson polynomials as a representation space of some
quadratic algebra.

The Sklyanin algebra possesses a representation by finite-difference operators with elliptic
coefficients which depend on an arbitrary complex parameter $g$ labeling representations~\cite{skl2}.
Particular linear combinations of this algebra generators in a degeneration limit,
such that the elliptic nome goes to zero and the elliptic functions reduce to trigonometric ones,
take the form (the details of this procedure can be found in \cite{Gor93}, see also \cite{AA2008})
\begin{align}
&A(g)=  \frac{\textup{i}e^{\frac{\textup{i}\pi}{2} b^2}}{2} \frac{e^{-\pi \textup{i} b g}}{\sin 2\pi \textup{i} b z}
\left[ e^{2 \pi b z} e^{\frac{\textup{i} b}{2}\dd_z} - e^{-2 \pi b z} e^{-\frac{\textup{i} b}{2}\dd_z} \right] ,\lb{rep1} \\
&B(g)= -\frac{1}{2}\cos \pi b^2 \, C(g)  -\frac{1}{4\sin \pi b^2} \frac{1}{\sin 2 \pi \textup{i} b z} \times \notag \\
& \kern 32pt
\left[ \cos\pi b(2g + 4 \textup{i} z - b) \,e^{\frac{\textup{i} b}{2}\dd_z}
- \cos\pi b(2g - 4 \textup{i} z - b) \,e^{-\frac{\textup{i} b}{2}\dd_z} \right], \lb{rep2} \\
&C(g)= \frac{1}{2\sin \pi b^2} \frac{1}{\sin 2\pi \textup{i} b z}
\left[ e^{\frac{\textup{i} b}{2}\dd_z} - e^{-\frac{\textup{i} b}{2}\dd_z} \right] , \lb{rep3} \\
&D(g)= - \frac{\textup{i}e^{-\frac{\textup{i}\pi}{2} b^2}}{2} \frac{e^{\pi \textup{i} b g}}{\sin 2\pi \textup{i} b z}
\left[ e^{-2 \pi b z} e^{\frac{\textup{i} b}{2}\dd_z} - e^{2 \pi b z} e^{-\frac{\textup{i} b}{2}\dd_z} \right]. \lb{rep4}
\end{align}
These operators satisfy defining relations \eqref{alg}.
In this representation the Casimir operators \p{Cas} take the values
$$
K_0(g) = e^{\textup{i} \pi b^2} ,\qquad K_1(g) = \frac{\cos 2\pi b g}{2\sin^2 \pi b^2}.
$$

We can construct Verma module representations of the algebra \p{alg} following an
analogy with the $s\ell_2$ algebra.
We choose $|0 \rangle = 1$ to be a lowest weight vector in the representation annihilated by the
lowering operator $C$, $C |0 \rangle  = 0$. In order to obtain the basis of the Verma module
we act by the raising operator $B$ on the lowest weight vector a number of times:
$| k \rangle := B^k \,| 0 \rangle,\, k\in \mathbb{Z}_{\geq 0}$.
Using relations \p{alg} one can check that
\begin{align}
& A \, | k \rangle = \sum_{ l = 0}^{[k/2]} a_{k,l}(g) | k - 2 l \rangle \,,\;\;
D \, | k \rangle = \sum_{ l = 0}^{[k/2]} d_{k,l}(g) | k - 2 l \rangle \,,  \notag\\
& C \, | k \rangle = \sum_{ l = 0}^{[\frac{k-1}{2}]} c_{k,l}(g) | k - 1 - 2 l \rangle \,, \label{Verma}
\end{align}
where $a_{k,l},\,d_{k,l},\,c_{k,l}$ are some functions of $g$.
Contrary to the familiar situation of $s\ell_2$ the lowering operator
$C$ acting on the vector $| k \rangle$ produces not $| k-1 \rangle$ but a linear combination of
vectors with descending weights $k-1, k-3, k-5,\ldots$. Similarly, the operators $A$ and $D$ are not diagonal
contrary to their counterpart in $s\ell_2$. Acting on the vector $|k \rangle$ they mix it
with the vectors having descending weights $k-2, k-4, k-6, \ldots$.
For the chosen trigonometric polynomial realization, the vector $| k \rangle$ is
a linear combination of $\cos(2\pi \textup{i} j b z)$, where $j = k, k-2 ,k-4,\ldots, 1$ (or 0).

For the generic values of $g$ the representation is infinite-dimensional.
However, if $g = (n+1)\frac{b}{2},\, n \in \mathbb{Z}_{\geq 0},$ the situation drastically changes.
Then $| n+1 \rangle$ is a linear combination of $\{| k \rangle\}_{k=0}^{n}$. Acting
by powers of the raising operator $B$ on the vector $| n \rangle$ we do not get out of the $n$-dimensional space.
In order to avoid misunderstanding we note that $C\,| n+1 \rangle \neq 0$, unlike the $s\ell_2$ case.

Since representations with the labels $g$ and $-g$ have the same values of the Casimir
operators, they are equivalent. Indeed, they are intertwined by
the operator $\mathrm{M}(g)$, Eq.~\p{M}, as follows from the relations
\begin{align}\notag
&\mathrm{M}(g) \,A(g) = A(-g)\, \mathrm{M}(g),\;\;
\mathrm{M}(g) \,B(g) = B(-g)\, \mathrm{M}(g),\\
&\mathrm{M}(g) \,C(g) = C(-g)\, \mathrm{M}(g),\;\;
\mathrm{M}(g) \,D(g) = D(-g)\, \mathrm{M}(g),
\lb{intw1}\end{align}
which can be checked by an explicit calculation.

The operator $\mathrm{M}(g)$ is invariant under the swap $b \leftrightarrows b^{-1}$.
Therefore it is natural to introduce the second set of generators $\widetilde{A},\widetilde{B},\widetilde{C},\widetilde{D}$
respecting the commutation relations \p{alg} with the replacement $b \to b^{-1}$, i.e.
\begin{align}
&\widetilde{C} \,\widetilde{A} = e^{\frac{\textup{i} \pi}{b^2}} \,\widetilde{A}\, \widetilde{C} \,,\;\;
\widetilde{D} \,\widetilde{C} = e^{\frac{\textup{i} \pi}{b^2}} \,\widetilde{C}\, \widetilde{D} \,, \notag \\ &
[\,\widetilde{A}\,,\,\widetilde{D}\,] = -2 \textup{i} \sin^3 \frac{\pi}{b^2} \, \widetilde{C}^2 \,, \;\;
[\,\widetilde{B}\,,\,\widetilde{C}\,] = \frac{\widetilde{A}^2-\widetilde{D}^2}{2 \textup{i} \sin \frac{\pi}{b^2}} \,, \notag\\ &
\widetilde{A}\,\widetilde{B} - e^{\frac{\textup{i} \pi}{b^2}}\, \widetilde{B}\,\widetilde{A}
= e^{\frac{\textup{i} \pi}{b^2}} \, \widetilde{D} \, \widetilde{B} -\widetilde{B}\,\widetilde{D}
= \frac{\textup{i}}{2} \sin \frac{2 \pi}{b^2} \, (\widetilde{C}\,\widetilde{A}-\widetilde{D}\,\widetilde{C}).\lb{algtilde}
\end{align}
We also need to specify the commutation relations for generators from different sets.
The generators $A,D$ anticommute with $\widetilde{B},\widetilde{C}$;
the generators $B,C$ anticommute with $\widetilde{A},\widetilde{D}$;
the generators $A,D$ commute with $\widetilde{A},\widetilde{D}$;
and the generators $B,C$ commute with $\widetilde{B},\widetilde{C}$.

An explicit realization of the generators $\widetilde{A},\widetilde{B},
\widetilde{C},\widetilde{D}$ by finite-diffe\-rence operators is given by the formulae
\p{rep1}--\p{rep4}, where $b$ should be replaced by $b^{-1}$.
New Casimir operators have the form
\begin{align}
&\widetilde{K}_0 =  e^{\textup{i} \pi/b^2} \widetilde{A}\, \widetilde{D}
- \sin^2 \pi/b^2  \, \widetilde{C}^2 = e^{\textup{i} \pi/b^2} ,
\notag\\
&\widetilde{K}_1 = \frac{e^{-\textup{i} \pi/ b^2} \widetilde{A}^2
+ e^{\textup{i} \pi/b^2}\widetilde{D}^2}{4\sin^2 \pi/ b^2}
-\widetilde{B}\, \widetilde{C} -\frac{1}{2}\cos \pi/b^2 \, \widetilde{C}^2
= \frac{\cos 2\pi g/b}{2\sin^2 \pi/b^2}.
\label{Cas2}
\end{align}

Taken together, two sets of generators $A,B,C,D$
and $\widetilde{A},\widetilde{B},\widetilde{C},\widetilde{D}$ form the
algebra which we call {\it the hyperbolic modular double}.
It lies in between the Faddeev's modular double of $U_q(s\ell_2)$ \cite{fad:mod} and the elliptic modular double \cite{AA2008}
in the sense that these algebraic structures are related by a sequence of contractions
\begin{align*}
&\text{Elliptic modular double} \to \\ & \kern 4em \to \text{Hyperbolic modular double} \to \\ & \kern 9em \to \text{Modular double of $U_q(s\ell_2)$}.
\end{align*}
Using particular combinations of the generators of this algebra similar to the one
considered in \cite{Gor93}, it is possible to construct
a modular double of the Zhedanov algebra \cite{Z} as well.

Evidently, $\mathrm{M}(g)$ works as the intertwining operator for the second set of
generators as well,
\begin{align}\notag
\mathrm{M}(g) \,\widetilde{A}(g) = \widetilde{A}(-g)\, \mathrm{M}(g),\;\;
\mathrm{M}(g) \,\widetilde{B}(g) = \widetilde{B}(-g)\, \mathrm{M}(g),\\
\mathrm{M}(g) \,\widetilde{C}(g) = \widetilde{C}(-g)\, \mathrm{M}(g),\;\;
\mathrm{M}(g) \,\widetilde{D}(g) = \widetilde{D}(-g)\, \mathrm{M}(g).
\lb{intw2}\end{align}
Irreducible representations of the hyperbolic modular double are
fixed  by one complex number $g$, with the $g$ and $-g$ label representations
being equivalent. Realization of the generators of this algebra in the space
of analytical functions is unique (up to a multiplication by a numerical factor),
because solutions of a system of finite-difference equations with the shifts by $b$
and $b^{-1}$ (which can be taken real and incommensurate) are determined up to
multiplication by a number. Relations \eqref{intw1} and \eqref{intw2} are natural
extensions of the intertwining relations for the $U_q(sl_2)$ algebra and its
modular double derived by Ponsot and Teschner \cite{PT}. In the limit described
in \cite{spi:conm}, when the hyperbolic R-matrix is degenerated to that of
Faddeev and Volkov, the integral operator $\mathrm{M}(g)$ passes to
the intertwining operator of  \cite{PT}.

Intertwining operators are quite useful, since they enable one to get insight
to the structure of representations of the corresponding algebra.
Indeed, the null-space of $\mathrm{M}(g)$, $\mathrm{Ker}\,\mathrm{M}(g)$, and the image of $\mathrm{M}(-g)$, $\mathrm{Im}\,\mathrm{M}(-g)$,
are invariant spaces of the representation with the label $g$ that follows from \p{intw1} and \p{intw2}.

The inversion formula \p{sq} implies that
$$
\mathrm{M}_{z}(g) \gamma(g \pm \textup{i} x \pm \textup{i} z) = 0 ,\; \text{where} \; g = \frac{n b}{2} + \frac{m}{2b}
,\; n, m \in \mathbb{Z}_{\geq 0},\; (n,m) \neq (0,0),
$$
since the normalization factor $1/\gamma(2g)$ of $\mathrm{M}(-g)$ is divergent at
the specified values of $g$ (see Eq.~\p{zero}). Here the subindex in the operator
$\mathrm{M}_{z}$ indicates that $z$ is used as the integration variable.
Hence, expanding $\gamma(g \pm \textup{i} x \pm \textup{i} z)$ in $x$ we recover the null-space of $\mathrm{M}_{z}(g)$.
Moreover, the function $\gamma(g \pm \textup{i} x \pm \textup{i} z)$ is proportional to the
integrand of the operator $\mathrm{M}(-g)$.
Consequently, its expansion in $x$ lies in the image of $\mathrm{M}(-g)$.
Let us remind that for these values of $g$ the integral operator
$\mathrm{M}(g)$ turns to the finite-difference operator \p{Mfact}.

In the following we will be interested in irreducible finite-dimensional representations of the hyperbolic modular double at
\begin{align} \lb{g-lat}
g_{n,m} = \frac{b}{2} (n+1) + \frac{1}{2b} (m+1),\;\; n,m \in\mathbb{Z}_{\geq 0},
\end{align}
which have the dimension $(n+1)(m+1)$. They are realized in the invariant space
$$\mathrm{Ker}\,\mathrm{M}(g_{n,m}) \cap\mathrm{Im}\,\mathrm{M}_{ren}(-g_{n,m}),$$
where $\mathrm{M}_{ren}(-g)=\gamma(2g)\mathrm{M}(-g)$.

All basis vectors of the finite-dimensional irreducible representation are embraced by the generating function
$$
\textstyle \gamma(\pm \textup{i} x \pm \textup{i} z + g_{n,m}),
$$
where $x$ is an auxiliary parameter.
Indeed, owing to Eqs.~\p{period} and \p{refl}, it turns into the finite product of
trigonometric functions
\begin{align} \lb{gen-fun}
&\gamma(\pm \textup{i} x \pm \textup{i} z + g_{n,m})= \\ & \kern 4em
\prod_{r = 0}^{n-1}2\sin \pi b \textstyle (\pm \textup{i} x + \textup{i} z + \frac{b}{2}(n-1-2r)+\frac{1}{2b}(m+1)) \times \notag\\
& \kern 4em  \prod_{s = 0}^{m-1}2\sin \textstyle \frac{\pi}{b}(\pm \textup{i} x + \textup{i} z + \frac{1}{2b}(m-1-2s)-\frac{b}{2}(n-1)). \notag
\end{align}
From the latter formula we extract the natural basis of the finite-dimen-sional representation
$$
\cos(2j\pi \textup{i}  b z) \cos (2l \pi\textup{i}z/b) ,\;\; j = 0 ,1, \ldots, n ,
\;\; l =0 ,1, \ldots, m .
$$
Note that the generating function coincides with the edge Boltzmann weight \p{BW}.

\section{Reductions of the integral R-operator}
\lb{redR}

\begin{figure}[h] \begin{center}\includegraphics[width = 10 cm]{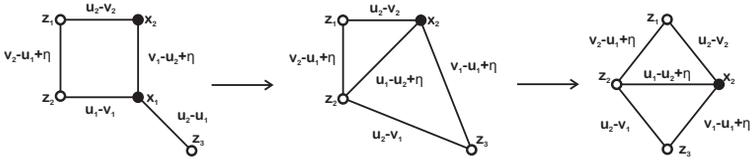}\end{center}
\caption{\label{fig4}The sequence of transformations converting the integral operator
$\mathbb{R}_{12}(u|g_1,g_2)$ at $g_1 = g_{n,m}$ to a finite-dimensional matrix in the first space.} \end{figure}

In this section we show that the integral operator solution, Eq.~\p{Rint}, of the YBE~\p{YBE}
enables one to recover all finite-dimensional solutions of YBE as well.
In order to do it we apply the operator $\mathbb{R}_{12}(u|g_1,g_2)$ to the function
$\gamma(\pm \textup{i} z_1 \pm \textup{i} z_3 + u_1-u_2)\Phi(z_2)$, where $z_3$ is
an auxiliary parameter and $\Phi(z_2)$ is an arbitrary function from the second space.
For $g_1 = g_{n,m}$ the first factor turns to the generating function, Eq.~\p{gen-fun}.
Temporarily we assume $g_1$ to be generic.
Computation of the result of this action is pictorially presented in Fig.~\ref{fig4},
where we slightly changed the graphical rules.
Now all edges represent the Boltzmann weights $\overline{W}$, Eq.~\p{BW}.
The black blob corresponds to the vertex Boltzmann weight $\rho$.
We omit rapidity lines and indicate corresponding differences of the rapidities explicitly.

At the first step we apply the star-triangle relation, Eq.~\p{HSTR}, implementing
integration at the vertex $x_1$. Thus the only integration left is at the vertex
$x_2$. It corresponds to the integral operator $\mathrm{M}_1(u_2-v_2)$
with the kernel $\rho(x_2)\,\overline{W}(u_2-v_2;z_1,x_2)$ which acts on the product
$\overline{W}(v_1-u_1+\eta;z_1,z_3)\,\overline{W}(u_1-u_2+\eta;z_1,z_2)\,\Phi(z_1)$.
At the second step we just rearrange the factors such that we gain the integral operator $\mathrm{M}_2(u_1-u_2+\eta)$
with the kernel $\rho(x_2)\,\overline{W}(u_1-u_2+\eta;z_2,x_2)$
which acts on the product $\overline{W}(u_1-v_2;z_1,z_2)\,\overline{W}(v_1-u_1+\eta;z_2,z_3)\,\Phi(z_2)$.
Then we note that the remaining integral operator
$\mathrm{M}_2(u_1-u_2+\eta) = \mathrm{M}_2(g_1+\eta)$ for $g_1 = g_{n,m}$
turns to the finite-difference operator
$\mathrm{M}_2({\textstyle  \frac{nb}{2}+\frac{m}{2b}})$, Eq.~\p{Mfact}.
Thus we have obtained the reduction formula which encompasses all solutions of the YBE~\p{YBE}
that have the symmetry of the hyperbolic modular double and
are realized on the tensor product of the finite-dimensional representation
with the label $g_{n,m}$, Eq.~\p{g-lat}, in the first space, and arbitrary infinite-dimensional
representation with the label $g$ in the second one,
\begin{align}\lb{reduct}
&\textstyle \mathbb{R}_{12}(u|g_{n,m},g) \,\gamma(\pm \textup{i} z_1 \pm \textup{i}
z_3 + \frac{b}{2} (n+1)+ \frac{1}{2b}(m+1))\,\Phi(z_2) \\ &
= c \cdot \frac{\gamma(\pm \textup{i} z_2 \pm \textup{i} z_3 + \frac{-u+g_{n,m}+g}{2})}
{\gamma(\pm \textup{i} z_1 \pm \textup{i} z_3 + \frac{-u-g_{n,m}-g-2\eta}{2})}\times \notag \\ & \kern 4em
\mathrm{M}_2({\textstyle  \frac{nb}{2}+\frac{m}{2b}})\,\frac{\gamma(\pm \textup{i} z_1 \pm \textup{i} z_2 + \frac{-u+g_{n,m}-g}{2})}
{\gamma(\pm \textup{i} z_2 \pm \textup{i} z_3 + \frac{-u-g_{n,m}+g-2\eta}{2})}\notag
\,\Phi(z_2),
\end{align}
where
$$
c=\frac{1}{\gamma(u+g_{n,m}\pm g)}.
$$
The same result can be obtained using the fusion following the procedure
described in \cite{CDS1,CDS}.

Expanding both sides of this formula in the auxiliary parameter $z_3$ we recover
the reduced R-operator that is a matrix whose entries are some
finite-difference operators acting in the second space, i.e. we have the L-operator.
One can straightforwardly reduce further the L-operator \eqref{reduct} to R-matrices
which are finite-dimensional in both spaces. In order to achieve it we just need to
force the representation
label $g$ of the second space to lie on the second copy of the lattice \p{g-lat}.

Remarkably, the factorized form of the integrand function of the integral operator
\p{Rint} is inherited by the reduced R-operator.
We will see in Sect.~\ref{factL} and \ref{allFact} that the reduced
solution of YBE \p{reduct} can be further arranged to the factorized product
of matrices form.
In \cite{CDS1,CDS} an analogous reduction formula has been derived for the integral
R-operators in the following three cases: 1) the R-operator with the symmetry group
$\mathrm{SL}(2,\mathbb{C})$; 2) solutions of YBE with the symmetry of the modular
double of $U_q(s\ell_2)$; 3) the most general known R-operator obeying
the symmetry of the elliptic modular double (and of the Sklyanin algebra, of course).

\section{The fundamental representation $\mL$-operator and its factorization}
\label{factL}
Let us show how formula \p{reduct} works in practice.
We consider the simplest nontrivial representation in the first space
$g_1 = g_{1,0} = b+\frac{1}{2b}$ (see Eq.~\p{g-lat}), i.e. the fundamental
representation of the $(A,B,C,D)$-generated part of the hyperbolic double and trivial
representation for the $(\widetilde{A},\widetilde{B},\widetilde{C},\widetilde{D})$-part.
Corresponding solution of the YBE is known as the (spin $\frac{1}{2}$) L-operator.
The generating function in this case has the following form (see Eq.~\p{gen-fun})
$$
\textstyle
\gamma(\pm \textup{i} z_1 \pm \textup{i} z_3 + b + \frac{1}{2b}) = 2 \cos 2\pi \textup{i} b z_1 + 2 \cos 2\pi \textup{i}  b z_3 =
\mathbf{e}_1 \, 2 \cos 2\pi \textup{i}  b z_3 + \mathbf{e}_2,
$$
where the basis of the 2-dimensional representation in the first space $\mathbb{C}^2$ is formed by
$\mathbf{e}_1 = 1$ and $\mathbf{e}_2=2\cos 2 \pi \textup{i} b z_1$.
The intertwining operator from \p{reduct} simplifies to
$\mathrm{M}_2(\frac{b}{2}) = c \cdot \frac{1}{\sin 2 \pi \textup{i} b z_2}
\left( e^{\frac{\textup{i}b}{2}\dd_2} - e^{-\frac{\textup{i}b}{2}\dd_2} \right)$ (see Eq.~\p{Mfact}).
To simplify the formulae we shift the spectral parameter $u \to u+\frac{b}{2}$.
Now we wish to rewrite formula \p{reduct} in a matrix form.
In the formula \p{reduct} we pull the hyperbolic gamma functions depending
on $\pm \textup{i} z_1 \pm \textup{i} z_2$ to the left and the hyperbolic gamma functions
depending on $\pm \textup{i} z_2 \pm \textup{i} z_3$ to the right.
Then we simplify them by means of Eqs.~\p{period} and \p{refl}.
Thus the right-hand side expression in \p{reduct} takes the form
\begin{align}
&\Bigl[ 2\cos 2 \pi \textup{i} b z_1 - 2\cos\pi b ( 2 \textup{i} z_2 + u + g) \Bigr] \times \notag \\
&\kern 4em e^{\frac{\textup{i}b}{2}\dd_2}
\Bigl[ 2\cos 2 \pi \textup{i} b z_3 - 2\cos\pi b ( 2 \textup{i} z_2 - u + g) \Bigr] \notag \\
&- \Bigl[ 2\cos 2 \pi \textup{i} b z_1 - 2\cos\pi b ( 2 \textup{i} z_2 - u - g) \Bigr] \times \notag \\
&\kern 3.5em e^{-\frac{\textup{i}b}{2}\dd_2}
\Bigl[ 2\cos 2 \pi \textup{i} b z_3 - 2\cos\pi b ( 2 \textup{i} z_2 + u - g) \Bigr].
\end{align}
Now it is straightforward to rewrite the reduced R-operator
$\mathbb{R}_{12}(u+\frac{b}{2}|g_{1,0},g) =: \mathrm{L}(u|g)$
in a matrix form in the basis $\{\mathbf{e}_1,\mathbf{e}_2 \}$ of $\mathbb{C}^2$,
using the definition of matrix elements $\mathrm{L}(u|g)\, \mathbf{e}_k :=
\sum_{i}\mathbf{e}_i \left[\mathrm{L}(u|g)\right]_{i,k}$,
\begin{align}
&\mathrm{L}(u|g) = \frac{1}{\sin 2 \pi \textup{i} b z}
\begin{pmatrix}
-2 \cos 2 \pi b (\textup{i} z + u_1) & -2 \cos 2 \pi b (\textup{i} z - u_1) \notag\\
1 & 1
\end{pmatrix} \times \\ & \kern 70 pt \times
\begin{pmatrix}
e^{\frac{\textup{i}b}{2}\dd} & 0 \\
0 & -e^{-\frac{\textup{i}b}{2}\dd}
\end{pmatrix}
\begin{pmatrix}
1 & -2 \cos 2 \pi b (\textup{i} z - u_2) \\
1 & -2 \cos 2 \pi b (\textup{i} z + u_2)
\end{pmatrix}.\lb{Lax}
\end{align}
Here we substituted $z_2 \to z$.
The rapidities $u_1,u_2$ are defined as $u_1 = \frac{u+g}{2}$ and $u_2 = \frac{u-g}{2}$ (recall Eq.~\p{uv}).
We stress that the $\mL$-operator is automatically obtained in the factorized form,
Eq.~\p{Lax}. This might be expected since the initial formula \p{reduct} obeys
similar factorization.

Choosing in the $\mL$-operator the second space representation label as
$g =g_{1,0} = b+\frac{1}{2b}$, i.e. restricting it to the fundamental representation
as well, we recover a $4 \times 4$ matrix solution of the YBE~\p{YBE}.
This solution differs from the standard trigonometric R-matrix with 6 nonzero entries~\cite{KS79}.
It is the R-matrix of the 7-vertex model \cite{Ant97,ADHR92}.

This factorized representation is analogous to the one found in \cite{Ant97,H95}
(compare Eq.~\p{Lax} with the normal-ordered factorized $\mL$-operator Eq.~(2.20)
in \cite{Ant97}). There the lateral matrices are identified with the trigonometric
intertwining vectors that provide the vertex-face correspondence between the
7-vertex model and a trigonometric SOS model.

The $\mL$-operator, Eq.~\p{Lax}, can be written in terms of the
degenerate Sklyanin algebra generators as well,
\begin{align} \lb{genLax}
&\mathrm{L}(u|g) =
2\left(\begin{array}{c}
- e^{-\textup{i} \pi b u} \, A(g) - e^{\textup{i} \pi b u}\, D(g) \\
\sin \pi b^2 \,C(g)
\end{array}\right. \\ \notag
&\kern 3em \left.\begin{array}{c}
- 4 \sin \pi b^2 \, B(g) - 2 \sin \pi b^2 (\cos 2\pi b u + \cos \pi b^2) \,C(g) \\
e^{\textup{i} \pi b u} \, A(g) + e^{-\textup{i} \pi b u}\, D(g)
\end{array}\right).
\end{align}
In \cite{Ant97} this L-operator \p{genLax} has been identified
with the quantum L-operator for the 2-particle trigonometric Ruijsenaars model.

Taking $g_1 = g_{0,1} = \frac{b}{2}+\frac{1}{b}$ one recovers in the same way
the second $\mL$-operator $\widetilde{\mL}(u|g)$ whose entries are generators
of the second half of the hyperbolic modular double.

Thus we see that our construction is self-consistent. A particular reduction of
the integral R-operator results in the generators of the hyperbolic modular double.
Consequently the latter quantum algebra is indeed the symmetry algebra of the
integral R-operator \p{explR}.
The algebraic interpretation of the rapidities stated above \p{uv} is correct.
The infinite-dimensional spaces in the construction of the integral R-operator
are equipped with the structure of representations of the hyperbolic modular double.

Implementing the reduction condition $g_3 = g_{1,0}$ in the YBE~\p{YBE} we obtain
an $\mathrm{RLL}$-relation,
\begin{align}
& \mathbb{R}_{12}(u-v|g_1,g_2)\,\mathrm{L}_1(u|g_1)\,\mathrm{L}_2(v|g_2) = \notag \\ &
=\mathrm{L}_2(v|g_2)\,\mathrm{L}_1(u|g_1)\,\mathbb{R}_{12}(u-v|g_1,g_2). \label{RLL}
\end{align}
Here the integral R-operator \p{explR} acts in a pair of infinite-dimensional spaces with
the representation labels $g_1$ and $g_2$ and intertwines the matrix product of two
L-operators. The lower index (1 or 2) of the L-operator enumerates the
infinite-dimensional spaces where it acts nontrivially, i.e. the entries of
$\mathrm{L}_i$ (see Eq.~\p{Lax}) are some difference operators in the variable $z_i$.
The RLL-relation \p{RLL} can be rewritten in terms of the rapidities as well
in a full analogy with Eq.~\p{YBErap},
\begin{align}
&\mathbb{R}_{12}(u_1,u_2|v_1,v_2)\,\mathrm{L}_1(u_1,u_2)\,\mathrm{L}_2(v_1,v_2)= \notag \\
&=\mathrm{L}_2(v_1,v_2)\,\mathrm{L}_1(u_1,u_2)\,\mathbb{R}_{12}(u_1,u_2|v_1,v_2),\label{RLLrap}
\end{align}
where $\mathrm{L}(u_1,u_2) := \mL(u|g_1)$,
$\mathrm{L}(v_1,v_2) := \mL(v|g_2)$ (recall Eq.~\p{uv}).

Now we can give a natural interpretation of the operator $\mathrm{S}(u)$, Eq.~\p{S},
which is one of the factors of the R-operator \p{R4fact}.
It implements the permutation of rapidities $(u_1,u_2,v_1,v_2) \mapsto (u_1,v_1,u_2,v_2)$
in the matrix product of two L-operators, i.e.
$$
\mathrm{S}(u_2-v_1)\,\mL_1(u_1,u_2)\,\mL_2(v_1,v_2) = \mL_1(u_1,v_1)\,\mL_2(u_2,v_2)\,\mathrm{S}(u_2-v_1).
$$
This statement can be checked by a straightforward calculation.
On the other hand, the R-operator itself implements the permutation of
the rapidities $(u_1,u_2,v_1,v_2) \mapsto (v_1,v_2,u_1,u_2)$
in Eq.~\p{RLLrap}. For more details of such permutation of parameters
in various models, see \cite{DKK07,DM,DS}.

\section{Factorized finite-dimensional solutions of the YBE}
\lb{allFact}

In the previous section we have shown that the reduction formula \p{reduct}
produces the $\mL$-operator in the factorized form from the fundamental
representation in the first space for the integral R-operator. Now we are
going to demonstrate that the same pattern persists for all finite-dimensional
representations, i.e. we show that the higher-spin solutions of the YBE can be
factorized as well. Finite-dimensional representations of the hyperbolic modular
double naturally factorize to products of finite-dimensional representations of
its two halves. Therefore without loss of generality we can consider
nontrivial representations for only one of its halves.
Thus, we choose $g_1 = g_{n,0} = \frac{b}{2} (n+1) + \frac{1}{2b}$,
$n\in \mathbb{Z}_{\geq 0}$, (recall Eq.~\p{g-lat}) in the reduction formula \p{reduct}.

The generating function of the $(n+1)$-dimensional
representation of interest takes the form (recall Eq.~\p{gen-fun})
\begin{align} \lb{gen-fun2}
\gamma( \pm \textup{i} z \pm \textup{i} x + g_{n,0} )
= \prod_{r = 0}^{n-1} \Bigl[ 2\cos 2 \pi \textup{i} b z + 2\cos \pi b( 2 \textup{i} x + b(n-1-2r)) \Bigr] \notag \\
= \sum_{j = 1}^{n+1} \psi_{n+2-j}^{(n)}(x)\, \varphi_{j}^{(n)}(z)
= \sum_{j = 1}^{n+1} \varphi_{n+2-j}^{(n)}(x)\,\psi_{j}^{(n)}(z),
\end{align}
where
$$
\varphi^{(n)}_j (z) := (2\cos 2 \pi \textup{i} b z)^{j-1},\;\;\; j = 1, 2,\ldots,n+1.
$$
The second equality in \eqref{gen-fun2} is used to define the dual basis
$\psi^{(n)}(x)$, whereas the third equality follows from the invariance of the
generating function under the permutation of $x$ and $z$.
Thus the generating function produces two natural bases
$\{\mathbf{e}_{j}\}_{j=1}^{n+1}$ and $\{\mathbf{f}_j\}_{j=1}^{n+1}$ of $\mathbb{C}^{n+1}$,
$$
\mathbf{e}_{j} = \varphi^{(n)}_j(z)\;\; , \;\;
\mathbf{f}_{j} = \psi^{(n)}_j(z) \;\; , \;\; j = 1 , 2, \ldots , n+1.
$$
Expanding both sides of Eq.~\p{reduct} as linear combinations of $\varphi^{(n)}(z_3)$,
we obtain a matrix form of the reduced R-operator written in the indicated pair of bases,
i.e.
$$
\mathbb{R}_{12}(u|g_{n} ,\,g)\,\psi^{(n)}_{j}(z_1) =
\varphi^{(n)}_{l}(z_1) \bigl[\mathbb{R}_{12}(u|g_{n} ,\,g)\bigr]_{lj}.
$$
In the previous section we did not have such a subtlety, since for $n = 1$
(the fundamental representation) both bases coincide.

Similar to the pattern given in the previous section, the direct calculation yields
the following factorization formula for the reduced R-operator
\be \lb{factellip}
\mathbb{R}_{12}(u|g_{n,0} ,\,g) = V(u+g,z) \,D(z,\dd)\, \mathbf{C} \, V^{T}(u-g,z)
\, \mathbf{C},
\ee
consisting of the product of five matrices. Here we substituted $z_2 \to z$ for brevity.
$\mathbf{C}$ is a numerical matrix with the unities on the antidiagonal, i.e.
$\left( \mathbf{C} \right)_{lj} = \delta_{n+2-l , j}$.
Entries of the diagonal matrix
\begin{align}
\left[D(z,\dd)\right]_{lj}:= \delta_{lj}\, \beta^{(n)}_{l}(z)\,e^{(n+2-2l) \frac{\textup{i}b}{2}\dd_z}.
\end{align}
are the shift operators determined by the expansion of
$\mathrm{M}({\textstyle\frac{nb}{2}})$ of the form
\be \lb{Mintw}
\mathrm{M}({\textstyle\frac{nb}{2}}) = \sum_{l = 1}^{n+1} \beta^{(n)}_l(z) \,e^{(n+2-2l)\eta\dd_z}.
\ee
Entries of the matrix $V$, $\left[V(u,z)\right]_{jl} = V^{(n)}_{jl}(u,z)$, are
some trigonometric functions. They are defined by the relations
\begin{align}\label{Vdef}
&\sum_{j = 1}^{n+1} \varphi_j^{(n)}(x)\,V_{jl}^{(n)}(u,z) := \\ & \kern 4em
\prod_{r = 0}^{l-2} \Bigl[ 2\cos 2 \pi \textup{i} b x - 2\cos \pi b ( 2\textup{i} z - u - 2\eta - g_{n,0} + 2 b r) \Bigr] \times \notag \\
& \kern 4em  \prod_{r = 0}^{n-l} \Bigl[ 2\cos 2 \pi \textup{i} b x - 2\cos \pi b ( -2\textup{i} z - u - 2\eta - g_{n,0} + 2 b r) \Bigr]. \notag
\end{align}
It is easy to see that $V_{jl}^{(n)}(u,-z) = V_{j,n+2-l}^{(n)}(u,z)$, i.e.
$V(u,-z) = V(u,z)\,\mathbf{C}$.

Let us recall that for the factorized $\mL$-operator the lateral matrices are composed out of
the trigonometric intertwining vectors providing the vertex-face correspondence \cite{Ant97}.
Then, in the case of the $(n+1)$-dimensional representation, the lateral matrices $V$
are constructed out of the fused trigonometric intertwining vectors (see Eq.~\p{Vdef})
providing the vertex-face correspondence for the higher-spin models.
In \cite{CD15} an analogous factorization formula has been derived for
finite-dimensional R-operators with the symmetry algebras $s\ell_2$, $U_q(s\ell_2)$, and
the Sklyanin algebra.

\section*{Acknowledgements}

We are indebted to S. E. Derkachov for useful discussions of the results
of this paper. This work is supported by the Russian Science Foundation
(project no. 14-11-00598).



\begin{thebibliography}{99}


\bibitem{ADHR92}
  F.~C.~Alcaraz, M.~Droz, M.~Henkel, and V.~Rittenberg,
  Reaction-diffusion processes, critical dynamics and quantum chains,
  Annals Phys., {\bf 230} (1994), 250--302.

\bibitem{aar}
G. E. Andrews, R. Askey, and R. Roy,
Special Functions, Encyclopedia of Math. Appl.,
{\bf 71}, Cambridge Univ. Press, Cambridge, 1999.

\bibitem{Ant97}
A.~Antonov, K.~Hasegawa, and A.~Zabrodin,
On trigonometric intertwining vectors and nondynamical R matrix for
the Ruijsenaars model, Nucl. Phys. B, {\bf 503} (1997), 747--770.

\bibitem{bar} E. W. Barnes,
On the theory of the multiple gamma function, Trans. Cambridge
Phil. Soc., {\bf 19} (1904), 374--425.

\bibitem{Baxter}
R.~J.~Baxter, Exactly Solved Models in Statistical
Mechanics, Academic Press, London, 1982.

\bibitem{BMS}
V.~V.~Bazhanov, V.~V.~Mangazeev, and S.~M.~Sergeev,
Faddeev-Volkov solution of the Yang-Baxter equation and discrete conformal symmetry,
Nucl. Phys. B, {\bf 784} (2007), 234--258.

\bibitem{BS10}
  V.~V.~Bazhanov and S.~M.~Sergeev,
A master solution of the quantum Yang-Baxter equation and classical discrete integrable equations,
 Adv. Theor. Math. Phys., {\bf 16} (2012), 65--95.

\bibitem{Brad01}
 H.~W.~Braden, A.~Gorsky, A.~Odessky, and V.~Rubtsov,
Double elliptic dynamical systems from generalized Mukai-Sklyanin algebras,
Nucl. Phys. B, {\bf 633} (2002), 414--442.

\bibitem{BT06}
  A.~G.~Bytsko and J.~Teschner,
Quantization of models with non-compact quantum group symmetry:
Modular XXZ magnet and lattice sinh-Gordon model,
  J. Phys. A, {\bf 39} (2006), 12927.

\bibitem{CD14}
  D.~Chicherin and S.~Derkachov,
The R-operator for a modular double,
  J. Phys. A, {\bf 47} (2014), 115203.

\bibitem{CD15}
  D.~Chicherin and S.~Derkachov,
Matrix factorization for solutions of the Yang-Baxter equation,
    Zap. Nauchn. Semin. POMI, {\bf 433} (2015), 156--185.

\bibitem{CDKK12}
  D.~Chicherin, S.~Derkachov, D.~Karakhanyan, and R.~Kirschner,
 Baxter operators with deformed symmetry,
 Nucl. Phys. B, {\bf 868} (2013), 652--683.

\bibitem{CDS1}
D.~Chicherin, S.~E.~Derkachov, and V.~P.~Spiridonov,
From principal series to finite-dimensional solutions of the
Yang-Baxter equation, SIGMA, {\bf 12} (2016), 028.

\bibitem{CDS}
D. Chicherin, S.~E. Derkachov, and V.~P. Spiridonov,
New elliptic solutions of the Yang-Baxter equation,
Commun. Math. Phys., {\bf 345} (2016), 507--543.

\bibitem{SD}
S. E. Derkachov,  Factorization of the $R$-matrix. I.,
Zap. Nauchn. Sem. POMI, {\bf 335} (2006), 134--163;
J. Math. Sciences, {\bf 143} (2007), 2773--2790.

\bibitem{DKK01}
  S.~E.~Derkachov, D.~Karakhanyan, and R.~Kirschner,
  Universal R-matrix as integral operator,
  Nucl. Phys. B, {\bf 618} (2001), 589--616.

\bibitem{DKK07}
  S.~Derkachov, D.~Karakhanyan, and R.~Kirschner,
Yang-Baxter R operators and parameter permutations,
  Nucl. Phys. B, {\bf 785} (2007), 263--285.

\bibitem{DKM}
S. E. Derkachov, G. P. Korchemsky, and A. N. Manashov,
Noncompact Heisenberg spin magnets from high-energy QCD: 1.
Baxter $Q$ operator and separation of variables,
Nucl. Phys. B, {\bf 617} (2001), 375--440.

\bibitem{DM}
S.~Derkachov and A.~Manashov,
General solution of the Yang-Baxter equation with the symmetry group SL(n, $\mathbb{C}$),
Algebra i Analiz, {\bf 21} (2009), 1--94;
St. Petersburg Math. J., {\bf 21} (2010), 513--577.

\bibitem{DS} S.~E.~Derkachov and V.~P.~Spiridonov,
Yang-Baxter equation, parameter permutations, and the elliptic beta integral,
Uspekhi Mat. Nauk, {\bf 68} (2013), 59--106;
Russian Math. Surveys, {\bf 68}  (2013), 1027--1072.

\bibitem{DS2} S.~E.~Derkachov and V.~P.~Spiridonov,
Finite dimensional representations of the elliptic modular double,
Teor. Mat. Fiz., {\bf 183} (2015), 163--176; Theor. Math. Phys., {\bf 183} (2015), 597--618.

\bibitem{Fad94}
  L.~D.~Faddeev,
  Current-like variables in massive and massless integrable models,
  In: Quantum groups and their applications in physics, Amsterdam, 1996, (eds. L. Castellani, J. Wess), pp. 117--135.

\bibitem{F95}
  L.~D.~Faddeev,
  Discrete Heisenberg-Weyl group and modular group,
  Lett. Math. Phys., {\bf 34} (1995), 249--254.

\bibitem{Fad96}
  L.~D.~Faddeev,
  How algebraic Bethe ansatz works for integrable model,
  In: Quantum Symmetries, Proc. Les-Houches summer school, LXIV, Univ. Pr., Cambridge, 1997, (eds. J.A. Marck , J.P. Lasota),
    pp. 149--211.

\bibitem{fad:mod} L. D. Faddeev,
Modular double of a  quantum group,
Math. Phys. Stud., {\bf 21}, Kluwer, Dordrecht, 2000, 149--156.

\bibitem{FKV}
  L.~D.~Faddeev, R.~M.~Kashaev, and A.~Y.~Volkov,
Strongly coupled quantum discrete Liouville theory. 1. Algebraic approach and duality,
  Commun. Math. Phys., {\bf 219} (2001), 199--219.

\bibitem{Gor93}
  A.~S.~Gorsky and A.~V.~Zabrodin,
  Degenerations of Sklyanin algebra and Askey-Wilson polynomials,
  J. Phys. A, {\bf 26} (1993), 635--639.

\bibitem{H95}
K.~Hasegawa, Ruijsenaars commuting difference operators as commuting transfer matrices,
Commun. Math. Phys., {\bf 187} (1997), 289--325.

\bibitem{jm}
M. Jimbo and T. Miwa,
Quantum KZ equation with $|q|=1$ and correlation functions of the XXZ model in the gapless
regime,  J. Phys. A: Math. Gen., {\bf 29} (1996), 2923--2958.

\bibitem{Kul83}
  P.~P.~Kulish and N.~Y.~Reshetikhin,
  Quantum linear problem for the Sine-Gordon equation and higher representation,
  Zap. Nauchn. Semin. LOMI, {\bf 101} (1981), 101--110;
J. Sov. Math., {\bf 23} (1983), 2435--2441.

\bibitem{KS79}
  P.~P.~Kulish and E.~K.~Sklyanin,
  Quantum inverse scattering method and the Heisenberg ferromagnet,
  Phys. Lett. A, {\bf 70} (1979), 461--463.

\bibitem{KS81}
  P.~P.~Kulish and E.~K.~Sklyanin,
Quantum Spectral Transform Method. Recent Developments,
  Lect. Notes Phys., {\bf 151} (1982), 61--119.

\bibitem{MT}
    C. Meneghelli and  J. Teschner,
    Integrable light-cone lattice discretizations from the universal R-matrix,
arXiv:1504.04572.

\bibitem{NM} N. Noumi and K. Mimachi,
Rogers's $q$-ultraspherical polynomials on a quantum 2-sphere,
Duke Math. J., {\bf 63} (1991), 65--80.

\bibitem{PT}
B.~Ponsot and J.~Teschner,
Clebsch-Gordan and Racah-Wigner coefficients for a continuous series of
representations of $U_q(sl(2,R))$, Commun. Math. Phys., {\bf 224} (2001),
613--655.

\bibitem{Rah}
M.~Rahman, An integral representation of a ${}_{10}\phi_9$ and continuous bi-orthogonal ${}_{10}\phi_9$
rational functions, Can. J. Math., {\bf 38} (1986), 605--618.

\bibitem{Ru}
S.~N.~M.~Ruijsenaars, First order analytic difference equations and integrable quantum systems,
J. Math. Phys., {\bf 38} (1997), 1069--1146.

\bibitem{skl1} E. K. Sklyanin, Some algebraic structures connected with
the Yang-Baxter equation, Funkz. Analiz i ego Pril., {\bf 16} (1982), 27--34.

\bibitem{skl2} E. K. Sklyanin, On some algebraic structures
related to Yang-Baxter equation: representations of the quantum algebra,
Funkz. Analiz i ego Pril., {\bf 17} (1983), 34--48.

\bibitem{Smir10}
A.~Smirnov, Degenerate Sklyanin algebras,
Central Eur. J. Phys., {\bf 8} (2010), 542.

\bibitem{spi:umn}
V. P. Spiridonov,
On the elliptic beta function, Uspekhi Mat. Nauk, {\bf 56} (2001), 181--182;
Russian Math. Surveys, {\bf 56} (2001), 185--186.

\bibitem{spi:bailey}
V. P. Spiridonov, A Bailey tree for
integrals, Teor. Mat. Fiz., {\bf 139} (2004), 104--111; Theor. Math.
Phys., {\bf 139} (2004), 536--541.


\bibitem{AA2008}
V. P. Spiridonov,
Continuous biorthogonality of the elliptic hypergeometric function,
Algebra i Analiz, {\bf 20} (2008), 155--185;
St. Petersburg Math. J., {\bf 20} (2009), 791--812.

\bibitem{spi:conm} V. P. Spiridonov, Elliptic beta integrals
and solvable models of statistical mechanics, Contemp. Math.,
{\bf 563} (2012), 181--211.

\bibitem{SV2014}
 V.~P.~Spiridonov and G.~S.~Vartanov, Vanishing superconformal
indices and the chiral symmetry breaking,
J. High Energy Phys., {\bf 06} (2014), 062.

\bibitem{SW}
V. P. Spiridonov and S. O. Warnaar,
Inversions of integral operators and elliptic beta integrals on root
systems, Adv. Math., {\bf 207} (2006), 91--132.

\bibitem{St}
J.~V.~Stokman, Hyperbolic beta integrals, Adv. in Math.,
{\bf 190} (2004), 119--160.

\bibitem{Vol05}
  A.~Yu.~Volkov, Noncommutative hypergeometry,
  Commun. Math. Phys., {\bf 258} (2005), 257--273.

\bibitem{VF1}   A.~Yu.~Volkov and L.~D.~Faddeev,
Quantum inverse scattering method on a space-time lattice,
Teor. Mat. Fiz., {\bf 92} (1992), 207--214;
Theor. Math. Phys., {\bf 92} (1992), 837--842.

\bibitem{VF}
A.~Yu.~Volkov and L.~D.~Faddeev, Yang-Baxterization of the
quantum dilogarithm, Zap. Nauchn. Semin. POMI, {\bf 224} (1995), 146--154;
J. Math. Sciences, {\bf 88} (1998), 202--207.

\bibitem{Z}
A. S. Zhedanov, Hidden symmetry of Askey-Wilson polynomials,
Teor. Mat. Fiz., {\bf 89}  (1991), 190--204;
Theor. Math. Phys., {\bf 89}  (1991), 1146--1157.

\end{thebibliography}
\end{document}